\documentclass[reprint,amssymb,aps,pre,twocolumn,floats,showpacs]{revtex4-1}
\usepackage{epsfig}
\usepackage{graphicx}
\usepackage{color}
\usepackage{bm}
\usepackage{latexsym}
 \usepackage{epstopdf}

\usepackage{amsfonts,amsmath}
\usepackage{dcolumn}
\usepackage{latexsym}

\usepackage{color}
\usepackage{ulem}

\usepackage{natbib}

\begin{document}

\newcommand{\hide}[1]{}
\newcommand{\tbox}[1]{\mbox{\tiny #1}}
\newcommand{\half}{\mbox{\small $\frac{1}{2}$}}
\newcommand{\sinc}{\mbox{sinc}}
\newcommand{\const}{\mbox{const}}
\newcommand{\trc}{\mbox{trace}}
\newcommand{\intt}{\int\!\!\!\!\int }
\newcommand{\ointt}{\int\!\!\!\!\int\!\!\!\!\!\circ\ }
\newcommand{\eexp}{\mbox{e}^}
\newcommand{\bra}{\left\langle}
\newcommand{\ket}{\right\rangle}
\newcommand{\EPS} {\mbox{\LARGE $\epsilon$}}
\newcommand{\ar}{\mathsf r}
\newcommand{\im}{\mbox{Im}}
\newcommand{\re}{\mbox{Re}}
\newcommand{\bmsf}[1]{\bm{\mathsf{#1}}}
\newcommand{\mpg}[2][1.0\hsize]{\begin{minipage}[b]{#1}{#2}\end{minipage}}

\title{Statistics of coherent waves inside media with L\'evy disorder}

\author{Luis A. Razo-L\'opez$^{1}$, Azriel Z. Genack$^{2,3}$, Victor A. Gopar$^{4,\dagger}$}
\affiliation{
$^1$Universit\'e C\^ote d'Azur, CNRS, Institut de Physique de Nice, Parc Valrose. 06100 Nice, France.\\
$^{2}$Department of Physics, Queens College of the City University of New York, Flushing, NY, 11367 USA.\\
$^3$The Graduate Center of the City University of New York, New York, NY, 10016 USA\\
$^4$Departamento de F\'isica Te\'orica, Facultad de Ciencias, and BIFI,  Universidad de Zaragoza, Pedro Cerbuna 12, ES-50009 Zaragoza, Spain.}

\begin{abstract}

Structures with heavy-tailed distributions of disorder occur widely in nature.  The evolution of such systems, as in foraging for food or the occurrence of earthquakes is generally analyzed in terms of an incoherent series of events. But the study of wave propagation or lasing in such systems requires the consideration of coherent scattering. We consider the distribution of wave energy inside 1D random media in which the spacing between scatterers follow a L\'evy $\alpha$-stable distribution characterized by a power-law decay with exponent $\alpha$. We show that the averages of the intensity and logarithmic intensity are given in terms of the average of the logarithm of transmission and the depth into the sample raised to the power $\alpha$.  Mapping the depth into the sample to the number of scattering elements yields intensity statistics that are identical to those found for Anderson localization in standard random media. This allows for the separation for the impacts of disorder distribution  and wave coherence in random media.
\end{abstract}


\maketitle

\section{Introduction}

The average motion of particles in space and time in samples in which at least the first two moments of the distribution of spacing are finite follows a diffusion equation in Brownian models. Coherent-transport phenomena of classical and 
quantum waves have been also studied within Brownian approaches \cite{Beenakker-review,Mello_book}. Diffusion is suppressed as a result coherent backscattering in which waves returning to points in the medium along time-reversed paths interfere constructively. Anderson localization  occurs as diffusion ceases in sufficiently large systems of dimensions $d \le 2$ and in higher dimensions above a critical value of disorder \cite{Anderson,Abrahams1979,Abrahams2009,Kramer}. In 1D, all waves are localized. The average transmission falls exponentially asymptotically, $\langle T \rangle \sim \exp{(-L/\ell)}$, while $\langle \ln T \rangle =- L/\ell$. Indeed, the full statistics of transmission in standard light-tailed distributions of separation between scattering elements is determined in accord to the single parameter scaling  theory of localization in terms of the dimensionless parameter $L/\ell$~\cite{Anderson_1,Mello_book}. 

Most of studies of coherent transport  in random media consider standard light-tailed 
distributions of disorder  that lead to Anderson localization. Such systems include mesoscopic electronic systems in micron scale devices at low temperatures and classical waves in stationary media. However, heavy tailed distributions are common in biology and geology~\cite{Zaburdaev} and may lead to advantageous mesoscopic devices.

Heavy-tailed L\'evy $\alpha$-stable distribution are characterized by power-law tails. Thus,  for a random variable $z$ following a L\'evy $\alpha$-stable distribution $\rho(z)$~\cite{Levy,Kolmogorov,Uchaikin}:
\begin{equation}
\label{qalpha}
 \rho(z) \sim  1/z^{1+\alpha}
\end{equation}
for $z \gg 1$ 
and $0 < \alpha < 2$. For $\alpha < 1$, both the first and second moments diverge.

In L\'evy-type disorder, waves can travel long distances without being scattered and thus have a profound impact on the transport properties. 
Measurements and analytic calculations of wave transmission in L\'evy 
$\alpha$-stable media give different scaling than in standard 1D random media
~\cite{Lambert1998,Barthelemy2008, Mercadier2009, Falceto2010, Ilias2012,Fernandez2014,AntonioF2012,Asatryan2018,Barbosa2019, Razo2020,Boose, Beenakker2009,Burioni2010,Sibatov,Bouchad1990,Zaburdaev}.

In this work, we treat the energy inside heavy-tailed L\'evy disordered media. 
The distinctive impacts of L\'evy $\alpha$-stable disorder and Anderson localization upon wave propagation are manifest. Potential application of novel states in L\'evy disordered media for low threshold lasing are discussed.

In particular, we investigate the statistics of waves in-
side a 1D L\'evy disordered structures via the statistics of
intensity $I(x)$ at the observation point $x$ (See Fig. 1).

\begin{figure}
\includegraphics[width=0.95\columnwidth]{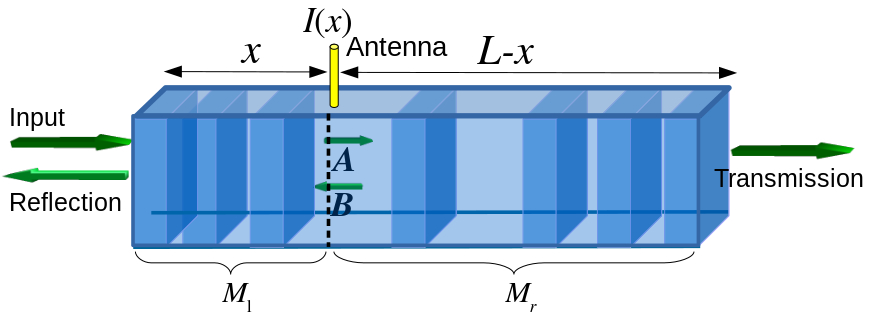}
\caption{Schematic of a random waveguide with scatterers (slabs) randomly separated according to a L\'evy distribution. The scattering processes at the left and right of the observation point $x$ are described by the transfer matrices $M_l$ and $M_r$, respectively.}
\label{fig1}
\end{figure}

As we show below, for L\'evy-type disorder, the average of the intensity and its logarithm follows a power-law dependence with the observation point. We will contrast these results for disordered systems with standard disorder, which have been studied 
experimentally and theoretically with random-matrix theory~\cite{Yamilov2014, Mello2015, Cheng}.  Calculations of intensity 
inside L\'evy disordered samples \cite{Xujun_arxiv} using the concept of leap-over to compute the density of scatterers gives different results from those presented in Fig.~\ref{fig4} (solid lines).

The intensity inside L\'evy disordered samples has been analyzed in \cite{Xujun_arxiv} 
using the concept of leap-over to calculate the density of scatterers,
but discrepancies exist between the calculated average intensity and numerical simulations.

The present manuscript is organized as follows. In Section II, we present general expressions for the intensity and transmission in a single sample in terms of transfer matrices. We then introduce some known results for the statistics of the transmission of standard disordered systems that will be contrasted with L\'evy disorder in the subsequent section. In Section III, L\'evy type disorder is introduced and compared to transmission in standard  systems. The results for the transmission are useful in the study of the statistics of the intensity inside the medium. The averages of the intensity and of the logarithm of intensity are given as functions of depth into the sample. Examples of the complete distribution of the logarithmic intensity are shown in Section III. A summary of the results and discussion are given in Section IV.

\section{Preliminaries}

\subsection{Transfer matrix: transmission and intensity}

Let us assume that we measure the intensity $I(x)$ at a point $x$. If $A$ and $B$ are the amplitudes of the forward and backward  waves going at this point (see Fig.~ \ref{fig1}), $I(x)$ is   given 
by
\begin{equation}
\label{I}
 I(x)=|A \exp{(ikx)}+ B \exp{(-ikx)}|^2 , 
\end{equation}
where $k$ is the wavenumber. We now introduce the transfer matrices 
$M_l$ and  $M_r$ associated with the segments of the sample at the left and right-hand side of the observation point, respectively: 
\begin{equation}
\label{mmatrices}
M_{l(r)}=  \begin{bmatrix}
    \gamma_{l(r)} & \beta_{l(r)} \\
      \beta_{l(r)}^*  &  \gamma_{l(r)}^*
\end{bmatrix} ,
\end{equation}
where $\gamma_{l(r)}$ and $\beta_{l(r)}$ are  complex numbers satisfying $|\gamma_{l(r)}|^2-|\beta_{l(r)}|^2=1$. The amplitudes $A$ and $B$ 
can be written in terms  of the transfer matrices, $M_{l(r)}$ and from Eq. (\ref{I}),  the intensity is given by 
\begin{eqnarray}
\label{Iofx}
 I(x)&=&\frac{1}{|\gamma_r|^2}\left|\gamma_l \gamma^{*}_r+\beta_l \beta^*_r  \right|^2
 \left|1-\frac{\beta^*_r}{\gamma^*_r} \exp{-2ikx} \right|^2  \nonumber \\
 &=& \frac{T}{T_r}\left|1-\frac{\beta^*_r}{\gamma^*_r} \exp{(-2ikx)} \right|^2 ,
\end{eqnarray}
where  $T=\left|\gamma_l \gamma^{*}_r+\beta_l \beta^*_r  \right|^2$ is the transmission coefficient of the entire sample and $T_r=1/|\gamma_r|^2$ is the transmission coefficient of the right segment.

The transfer matrices $M_{l(r)}$ are conveniently written in the polar representation as \cite{Mello_book} 
\begin{eqnarray}
\label{mpolar}
 M_{l (r)} = \begin{bmatrix}
   \sqrt{1+\lambda_{l(r)}}e^{i \theta_{l(r)}}   &  \sqrt{\lambda_{l(r)}}e^{i(2\mu_{l(r)}-\theta_{l(r)})} \\
    \sqrt{\lambda_{l(r)}}e^{-i(2\mu_{l(r)}-\theta_{l(r)})}  & \sqrt{1+\lambda_{l(r)}}e^{-i \theta_{l(r)}}  \nonumber \\
\end{bmatrix} , 
\end{eqnarray}
with phases $\theta_{l (r)}, \mu_{l (r)}$$\in$$[0,2\pi]$ and  $\lambda_{l(r)} \ge 0$. An advantage of using the polar representation is that the  radial variables $\lambda_{l(r)}$ are directly related to the transmission coefficients: $\lambda_{l(r)}=\left(1-T_{l(r)}\right)/T_{l(r)}$. Therefore Eq. (\ref{Iofx}) can be written as
\begin{equation}
\label{Iofx_polar}
 I(x)= \frac{T}{T_r}\left|1-\sqrt{1-T_r} \exp{(-2i(\mu_r-\theta_r+kx))} \right|^2 ,
\end{equation}
while the total transmission $T$ is given by
\begin{eqnarray}
 \frac{1}{T}&=&\frac{1}{T_r T_l}\Big( 2+T_r T_l -T_r-T_l  \nonumber \\
 &&+\left. 2\sqrt{(T_r-1)(T_l-1)\cos{2(\mu_l-\mu_r +\theta_r)}}\right) ,
\end{eqnarray}

\subsection{Statistics of the transmission in standard disordered 1D media}

Now that we have obtained analytical expressions for the intensity and transmission of a single sample in the previous section, we  consider an ensemble 
of  random samples.  In particular,  
we assume that the disordered structures composed of  
randomly separated weak scatterers or slabs. Thus, the intensity $I(x)$ is a random quantity. From Eq.~(\ref{Iofx_polar}),  the statistics of $I(x)$ depend on the statistical properties of 
the transmission and the angular variables $\theta_{l (r)}$ and $\mu_{l (r)}$.

Before considering the case of disordered samples with L\'evy disorder, we   introduce the distribution of the transmission for standard disorder. The statistics of transmission through standard disordered systems with light-tailed distributions have been 
extensively studied using random matrix theory\cite{Mello_book,Beenakker-review}. The 
distribution of the transmission $p_s (T)$ is given by \cite{Molina,Kleftogiannis2013}
\begin{equation}
\label{pofT}
 p_s(T)=C  \frac{\left[\mathrm{acosh}(1/\sqrt{T})\right]^{1/2}}{T^{3/2}(1-T)^{1/4}}e^{-s^{-1}\mathrm{acosh}^2(1/\sqrt{T})} ,
\end{equation}
 where $C$ is a normalization constant, $s=L/\ell$ with $L$ the length of the system and $\ell$ the mean free path. The complete distribution of  transmission is determined by the parameter $s$, which is  proportional to the number of scatterers $n$ in the sample with proportionality constant $a$: $s=an$ \cite{Mello_groups}.

For later comparisons with systems with L\'evy disorder, we point out the asymptotic exponential decay with $L$ of the average transmission in standard disordered systems \cite{Beenakker-review}: 
 \begin{equation}
 \label{averageT}
  \langle T \rangle \propto \exp{\left(-L/2\ell \right)}
 \end{equation}
and the linear behavior of the average of the logarithmic transmission
 \begin{equation}
 \label{averagelnT}
  \langle - \ln T \rangle \propto L
 \end{equation}
 
 To illustrate some statistical properties of the transmission of standard disordered systems, we show in Fig.~\ref{fig_2}(a)  the distribution of the logarithmic transmission $p_s(\ln T)$, which is obtained from Eq. (\ref{pofT})  and the linear behavior of $\langle - \ln T \rangle$ with $L$,   given by Eq. (\ref{averagelnT}).  The histogram and symbols in Fig.~\ref{fig_2} are obtained from numerical simulations as explained next.

 The numerical simulations performed  in this work  are based on the transfer matrix approach \cite{Markos2008,AntonioF2012}. The numerical model consists  
 of layers of thickness 2.5 mm with refraction index $n_2=1.1$, and reflection coefficient 0.007, randomly placed in a background of index of refraction $n_1=1$  with separations 
 following a Gaussian distribution for standard disorder and a  L\'evy $\alpha$-stable distribution for L\'evy disorder. 
 We have fixed the frequency at 1 THz in all the calculations. 
 The statistics is collected for $10^6$  realizations of the disorder.

\section{Statistics of the intensity inside 1D media with L\'evy disorder}

We utilize the model of L\'evy disordered media introduced in ~\cite{Falceto2010} 
with the asymptotic decay given in Eq. (\ref{qalpha}). 
We briefly summarize the main results of Ref.~\cite{Falceto2010} for the transmission that  will be useful for obtaining  the statistical properties of the intensity. We will consider the case $\alpha<1$, where the effects of L\'evy disorder on  transport are strong.

\begin{figure}
\includegraphics[width=\columnwidth]{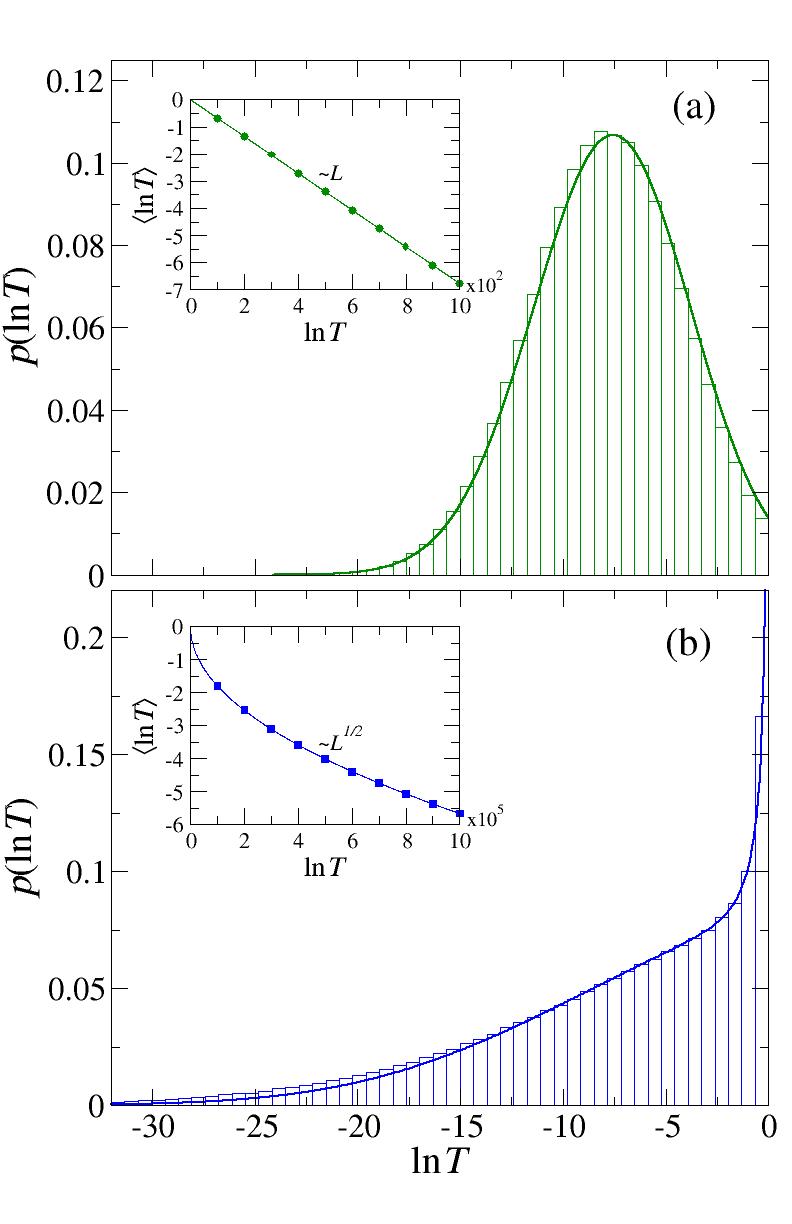}
\caption{(a) The distribution of the logarithmic transmission for standard disorder systems. The distribution is determined by the parameter $s=\langle -\ln T \rangle=8$. Inset: linear behavior of $\langle \ln T \rangle$ with the system length. (b) The distribution  
 of the logarithmic transmission for L\'evy  disordered systems. The distribution is determined by the parameters $\langle -\ln T \rangle=8$ and $\alpha=1/2$. Inset: the power-law behavior of $\langle \ln T \rangle$ with the system length, Eq.~ (\ref{lngofl_a}).}  
\label{fig_2}
\end{figure}

In  L\'evy disordered samples of fixed length $L$, the number of scattering units 
$n$ is a random variable with  strong sample-to-sample fluctuations; thus, it is crucial to know the complete distribution of $n$.  The probability density 
$\Pi_L(n;\alpha)$ of these fluctuations is given by \cite{Falceto2010}
\begin{equation}
\label{pi_alpha}
 \Pi_{L}(n;\alpha)=\frac2\alpha\frac{L}{(2n)^{\frac{1+\alpha}{\alpha}}}
q_{\alpha,c}\left({L}/{(2n)^{1/\alpha}}\right) ,
\end{equation}
for $0 < \alpha <1 $, in the limit $L\gg c^{1/\alpha}$ with $c$ a scaling parameter. The probability density $q_{\alpha,c}(x)$ has a power-law tail: $q_{\alpha,c}(x) \sim c/x^{1+\alpha}$ for  $x \gg 1$.

Using  the distribution of the transmission for a fixed number of scatterers  $p_{s}(T)$ given in Eq. (\ref{pofT}) and the distribution $\Pi_L(n,\alpha)$, Eq. (\ref{pi_alpha}), we write  the distribution of the transmission for 
L\'evy disordered systems $p_{\alpha,\xi}(T)$ as
\begin{eqnarray}
\label{pofT_alphaxi}
P_{\alpha,\xi}(T)=\int_0^\infty p_{s(\alpha,\xi,z)}(T) q_{\alpha,1}(z){\rm d}z ,
\end{eqnarray}
where $p_{s(\alpha,\xi,z)}(T)$ is given by Eq. (\ref{pofT}) with $s$ replaced by 
$s(\alpha,\xi,z)={\xi}/(2{z^\alpha \mathcal{I}_\alpha)}$ and $\mathcal{I_{\alpha}}$ is half of the  mean value 
$\langle z^{-\alpha} \rangle$: $\mathcal{I}_\alpha=(1/2)\int z^{-\alpha}q_{\alpha,1}dz=\cos(\pi\alpha/2)/2\Gamma(1+\alpha)$, where $\Gamma$ denotes the Gamma function \cite{Koren}. The parameter $\xi$ introduced in  Eq. (\ref{pofT_alphaxi}) is the average of the logarithmic transmission for a fixed length $L$: $\xi = \langle -\ln T \rangle_L = \int_0^{\infty} an \Pi_L(n)dn$, which is given by
\begin{equation}
\label{lngofl_a}
\langle \ln T\rangle_L = -\left( \frac{a}{c} \mathcal{I}_\alpha \right) L^\alpha,
\end{equation}
Since the factor in parentheses in Eq. (\ref{lngofl_a}) is a constant, 
the average of the logarithmic transmission is a power-law function of $L$, in contrast to the linear dependence for standard disorder in Eq.~(\ref{averagelnT}). Similarly, a power-law is found for the average transmission \cite{AntonioF2012} $\langle T \rangle \propto L^{-\alpha}$  in contrast to the exponentially decay for standard disordered systems in  Eq.~(\ref{averageT}).

In Fig.~\ref{fig_2}(b), we show an example of the distribution of the logarithm of transmission for L\'evy disordered structures characterized by $\alpha=1/2$. The theoretical results as given in Eq.~(\ref{pofT_alphaxi}) are compared to numerical simulations show as the histogram. The  power-law behavior 
of $\langle \ln T \rangle$ in L\'evy structures is shown in the inset 
in Fig. \ref{fig_2}(b). Thus, by comparing Fig.~ \ref{fig_2}(a) and \ref{fig_2}(b),  the 
strong influence of L\'evy disorder on the statistical properties of the transmission are clearly  seen.  
We also note that the only parameters that enter into Eq.~(\ref{pofT_alphaxi}) are $\alpha$ and 
$\xi=\langle \ln T \rangle$. Thus, the complete statistics of the transmission is determined by these parameters.

With the above results for the statistics of transmission, we now study the statistical properties of the intensity. 
As we show next, the presence of L\'evy disorder is revealed in basic  statistical quantities such as the ensemble averages $\langle \ln I(x) \rangle_L$ and $\langle I(x) \rangle_L$. 

Let us consider first the average of the logarithmic intensity, $\langle \ln I(x) \rangle$. The calculations are lenghty but  a simple analytical result can be provided. This quantity is of particular importance since it is directly related to  
the average of the logarithmic transmission, which along $\alpha$ determines  all the statistical properties  of the transport in L\'evy disordered systems.

We perform the average over the uniformly distributed phases in Eq.~(\ref{Iofx_polar}) to obtain 
$\langle \ln I(x) \rangle = \langle \ln T \rangle_L - \langle \ln T_r \rangle_{L-x}$. Since $\ln T$ is an additive quantity, we obtain $\langle \ln T \rangle_{L-x}=\langle \ln T \rangle_L - \langle \ln T \rangle_{x}$, and 
from Eq.~(\ref{pi_alpha}), we have 
\begin{eqnarray}
\label{logofIx}
 \langle \ln I(x) \rangle &=& - \left( \frac{a}{c} \mathcal{I}_\alpha \right)  x^\alpha \nonumber \\
 &=&  \langle \ln T \rangle_L \left( \frac{x}{L}\right)^\alpha .
\end{eqnarray}
Thus, $ \langle \ln I(x) \rangle \propto x^\alpha$ has a power-law behavior in L\'evy disordered media, in contrast 
to the linear dependence in standard disordered media~\cite{Cheng}.

\begin{figure}
\includegraphics[width=\columnwidth]{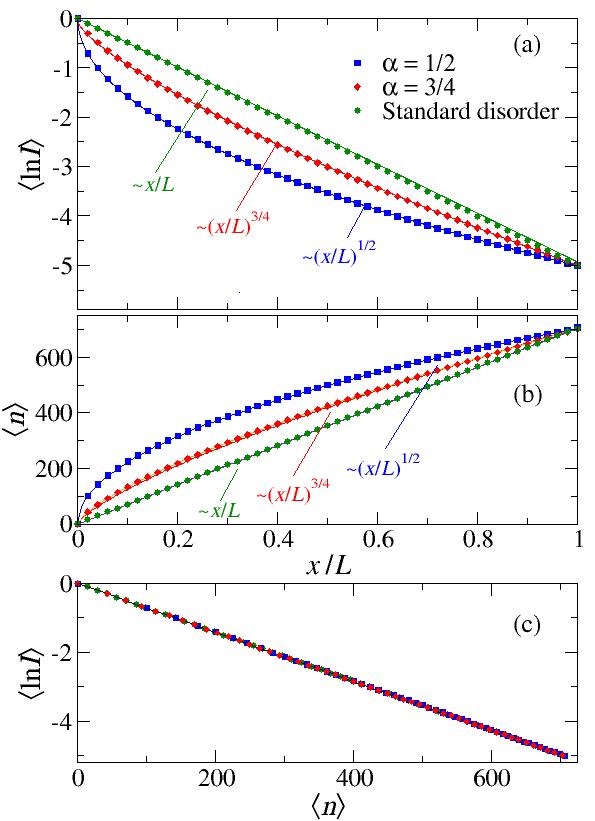}
\caption{(a) Average of the logarithmic intensity for L\'evy disordered systems with $\alpha=1/2$ and 3/4 (red and blue symbols). The solid lines are given by Eq. (\ref{logofIx}). For comparison, the linear dependence with $x$ in the case of standard disorder is also shown (green). In all cases, $\langle -\ln T \rangle_L=5$. (b) The average number of slabs $\langle n\rangle$ in the L\'evy ($\alpha=1/2, 3/4$) and standard disordered structures at the position of observation $x$. (c) $\langle \ln I \rangle$ as a function of $\langle n \rangle$ for $\alpha=1/2, 3/4$, and standard disorder.}
\label{fig3}
\end{figure}

We verify numerically the result given in Eq. (\ref{logofIx}). The results (symbols) are shown in Fig.~\ref{fig3}(a) together with the theoretical results (solid lines) from Eq.~(\ref{logofIx}) for $\alpha=1/2$ and 3/4. The linear behavior  of $ \langle \ln I(x) \rangle$  for standard  disorder is also shown in Fig.~\ref{fig3}(a) (green squares).
 
The $\langle \ln I(x) \rangle$ does not fall linearly in L\'evy structures as it does in standard (Fig.~\ref{fig3}(a)), however,   
the role of coherent backscattering in inhibiting propagation is unchanged. To gain 
an insight into this nonlinear behavior, we note that $\langle \ln I(x) \rangle$ is given by the difference between average of the logarithmic transmission of the complete sample ($L$) and the right segment  ($L-x$), as we have shown above.  
The power-law behavior of $\langle \ln I(x) \rangle$ finds its origin in the power-law of the variation of the number of scatterers up to the depth $x$. This is illustrated in Fig. \ref{fig3}(b), where the average number of scatterers $\langle n \rangle=\int n \Pi_x(n)dn=\mathcal{I_\alpha}x^\alpha/c$ is plotted at the position $x/L$ for L\'evy ($\alpha=1/2, 3/4$) and standard disordered structures. For L\'evy disordered samples, the average number of scatterers follows a power law  with  exponent $\alpha$. In contrast, in standard disorder, the average number of scatterers is a linear function with the system size. Thus, for both L\'evy and standard disorder, $\langle \ln I \rangle$ is a linear function of $\langle n \rangle$, as  shown in Fig.~\ref{fig3}(c) and $\langle \ln I(x) \rangle$ is additive, as in standard disordered structures \cite{Cheng,Yiming}.

We now study the  ensemble average intensity $\langle I(x) \rangle$,  after averaging $I(x)$, Eq. ~(\ref{Iofx}), over the uniformly distributed random phases, we  write $\langle I(x) \rangle$ as~\cite{Mello2015}
\begin{eqnarray}
\label{averageIx}
 && \langle I(x) \rangle = \nonumber \\  && \int_0^1 \int_0^1  \frac{T_l\left(2-T_r \right)}{T_l+T_r-T_lT_r} P_{\alpha,\xi_l}(T_l) P_{\alpha,\xi_r}(T_r) dT_l 
dT_r , \nonumber \\
\end{eqnarray}
where the distributions $P_{\alpha,\xi_l}(T_l)$ and $P_{\alpha,\xi_r}(T_r)$
are the distributions for the left and right segments, respectively, with 
$\xi_l =\langle \ln T_l \rangle= (x/L)^\alpha \langle \ln T \rangle_L$. For the right segment, $P_{\alpha,\xi_r}(T_r)$
$\xi_r =\langle \ln T_r \rangle= (1-(x/L)^\alpha) \langle \ln T \rangle_L$, according to Eq. (\ref{lngofl_a}). We can verify the particular cases at $x=0$ and $x=L$: for $x=0$, $T_l=1$ and $\langle I(0) \rangle=\langle (2-T_r) \rangle=2-\langle T \rangle$, while at $x=L$, $T_r=1$ and therefore 
$\langle I(L) \rangle= \langle T_r \rangle =\langle T \rangle$.

\begin{figure}
\includegraphics[width=\columnwidth]{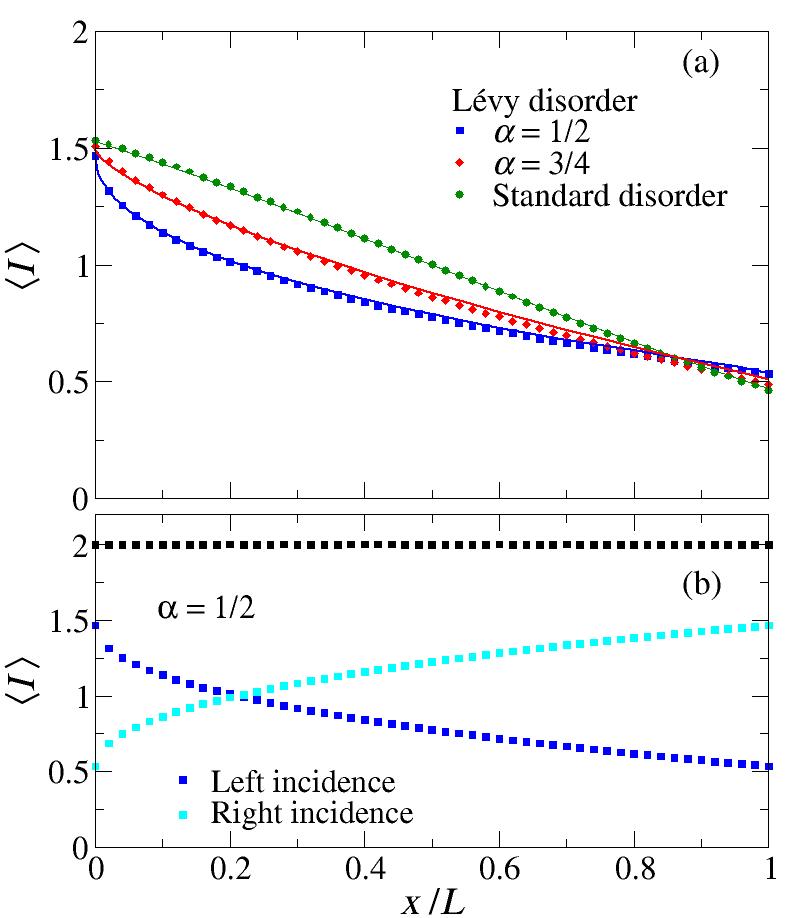}
\caption{(a) Ensemble average intensity $\langle I \rangle$ for L\'evy $\alpha=1/2$ and 3/4 and standard disorder. The solid lines are obtained from Eq.~(\ref{averageIx}). In all these 3 cases $\langle -\ln T \rangle=1$. (b) The average 
$\langle I \rangle$ for $\alpha=1/2$, as in (a), but for waves incident from the left and right incidences. The black squares show that the sum of both incidences is constant.}
\label{fig4}
\end{figure}

We perform numerical simulations to  support  Eq.~(\ref{averageIx}), where the double  integral is performed numerically. The distribution for the left segment  $P_{\alpha,\xi_l}(T_l)$ in Eq.~(\ref{averageIx}) is obtained from  Eq.~(\ref{pofT_alphaxi}), while   for the right segment,  
$P_{\alpha,\xi_r}(T_r)$ is obtained  by considering the corresponding 
probability density of scatterers which is generated numerically. 
The results are shown in Fig. \ref{fig4}(a) for $\alpha=1/2$ and 3/4. The numerical simulations and Eq.~(\ref{averageIx}) are seen in good agreement.  The average intensity for  standard disorder is also shown (green dots and solid line) in Fig \ref{fig4}(a) to provided a contrast with the power law dependence found in media with L\'evy disorder.

The profile of $\langle I(x) \rangle$ is not symmetric about the center, as it is in  standard homogeneously disordered systems. See Fig. \ref{fig4}(a).   
In  L\'evy disordered structures, the disorder is  inhomogeneous; the density of scatterers is greatest near the left side of the sample, as can be seen in Fig. \ref{fig3}(b),  where waves launched causing the intensity to fall more rapidly there.  
However, the sum of intensities for waves incident from the left and right is constant and equal to twice the intensity of the incident beam from one side, as shown in Fig. \ref{fig4}(b). This can be understood by noting that the integrand 
of Eq.~(\ref{averageIx}) for the wave incident from the right, $T_l$ and $\xi_l$ are replaced by 
$T_r$ and $\xi_r$, respectively, and similarly, $T_r$ and $\xi_r$ are replaced by $T_l$ and $\xi_l$ for waves incident from the right. Adding the contributions from the wave incident from the left and right gives $T_l(2-T_r)/(T_l+T_r-T_l T_r)+T_r(2-T_l)/(T_l+T_r-T_l T_r)=2$. Since $P_{\alpha,\xi_{l,(r)}}(T_{l,(r)})$ are normalized, the average of the sum of intensities excited by waves incident  on the left and right is equal to 2. This result is illustrated in Fig. \ref{fig4} (b) for the case of $\alpha=1/2$, where the black squares are the sum of the averages intensity for incident waves from the left and right ends of the samples.

The previous discussion is general and gives an average intensity profile for standard homogeneously disordered media which is symmetric about the center of the sample ~\cite{Yiming}. This symmetry is summarized by the expression  
$\langle I(x) \rangle + 
\langle I(L-x) \rangle =2$ and reflects the fact that the sum of intensity from the right and left is equal to the local density of states (LDOS) relative to LDOS outside the medium, which is unchanged by disorder.

\begin{figure}
\includegraphics[width=\columnwidth]{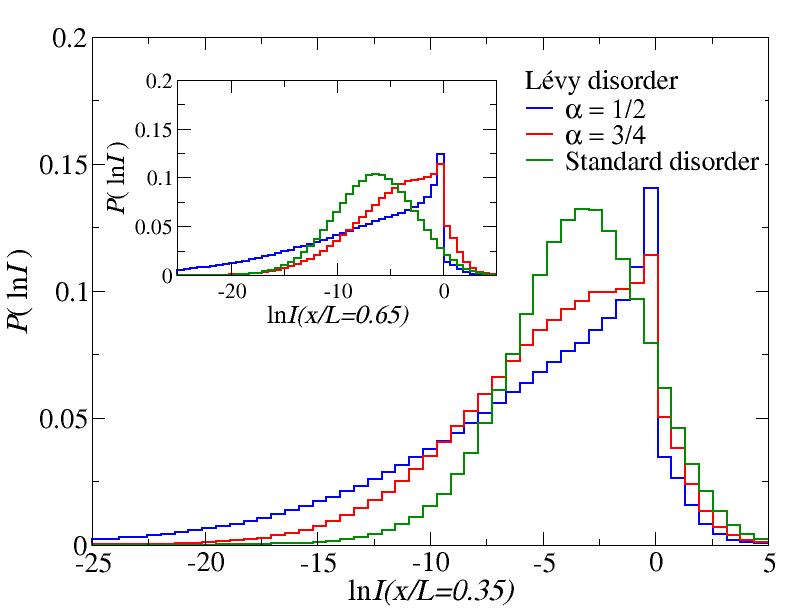}
\caption{Distribution of the logarithmic intensity for L\'evy and standard disordered structures. $\langle -\ln T \rangle=10$.}
\label{fig5}
\end{figure}

The complete distribution of the logarithmic is obtained numerically and shown in Fig.~\ref{fig5} for $\langle \ln T \rangle = 10$. There is a higher probability of large fluctuations of intensity in L\'evy disordered samples (blue and red histograms) as compared to standard disordered systems (green histogram).

\section{Summary and discussion}

We have studied  the wave intensity statistics inside  random 1D media with disorder described by L\'evy type distributions characterized by an asymptotic 
power-law decay. For both $\langle I(x) \rangle$ and $\langle \ln I(x) \rangle$, 
we find a power-law decay with position. 
In contrast, $\langle I(x) \rangle$  falls linearly in standard disordered systems.  
The slower decay with $x$ than for the standard disorder indicates that
wave localization in space is weaker in  L\'evy disorder than in standard disorder.

The equivalence of the statistics of intensity in  $\alpha$-stable L\'evy disordered systems  and standard random media at the corresponding layer number $n$, suggests opportunities for engineering structures for analyzing and controlling waves. The forward and background amplitudes within a layer of the medium are constant so that they interfere and create an oscillatory pattern with high peak intensity within the layer. When a thick layer is near the spatial and spectral peak of a mode, the lifetime of quasi-normal mode increases and the line narrows 
as the thickness of the layer increases. The material could therefore serve as a filter. If gain is introduced into this system, the correspondingly long lifetime of the mode would enhance the opportunity for emitted photons to stimulate emission before escaping the sample~\cite{Milner2005}. In addition, the large spatial extent of the mode allows the system to be efficiently pumped without saturating the gain medium.  
The prospects for an $\alpha$-stable laser will be considered in future work.

\begin{acknowledgments} 
We acknowledge discussions with Xujun Ma. 
 This work is supported by the National Science Foundation under EAGER Award No. 2022629 and by PSC-CUNY under Award No. 63822-00 51,  and by MCIU (Spain) under the Project number PGC2018-094684-B-C22. L.A.R.-L. thanks UCA$^{\rm{JEDI}}$ for the award of pre-doctoral studies fellowship. .
\end{acknowledgments}
$^\dagger$ Email: gopar@unizar.es

\bibliographystyle{unsrt}
\bibliography{Bibliography}

\end{document}